\documentclass[%
reprint,
superscriptaddress,
 amsmath,amssymb,
 aps,
 prx,
longbibliography,
floatfix,
]{revtex4-1}

\usepackage[utf8]{inputenc}
\usepackage{xcolor}
\usepackage[margin=0.76in]{geometry}
\usepackage{graphicx} 
\usepackage{physics}
\usepackage{amsmath}
\usepackage{siunitx}
\usepackage{multirow}
\usepackage{amssymb}
\usepackage{wasysym}
\usepackage{bibunits}
\usepackage{tabularx}
\usepackage{booktabs}
\usepackage{hyperref}
\hypersetup{
    colorlinks=true,
    linkcolor=blue,
    urlcolor=blue,  
    citecolor=blue,
}

\makeatletter
\def\maketitle{
	\@author@finish
	\title@column\titleblock@produce
	\suppressfloats[t]}
\makeatother

\usepackage[draft]{changes}

\begin{document}

\title{High threshold codes for neutral atom qubits with biased erasure errors}

\author{Kaavya Sahay}
\thanks{These authors contributed equally to this work.}
\affiliation{Yale University, Department of Applied Physics, New Haven, CT 06520, USA}
\affiliation{Yale Quantum Institute, Yale University, New Haven, Connecticut 06511, USA}
\author{Junlan Jin}
\thanks{These authors contributed equally to this work.}
\affiliation{Princeton University, Department of Electrical and Computer Engineering, Princeton, NJ, 08544, USA}
\author{Jahan Claes}
\affiliation{Yale University, Department of Applied Physics, New Haven, CT 06520, USA}
\affiliation{Yale Quantum Institute, Yale University, New Haven, Connecticut 06511, USA}
\author{Jeff D. Thompson}
\affiliation{Princeton University, Department of Electrical and Computer Engineering, Princeton, NJ, 08544, USA}
\author{Shruti Puri}
\affiliation{Yale University, Department of Applied Physics, New Haven, CT 06520, USA}
\affiliation{Yale Quantum Institute, Yale University, New Haven, Connecticut 06511, USA}

\date{\today}

\begin{abstract}
The requirements for fault-tolerant quantum error correction can be simplified by leveraging structure in the noise of the underlying hardware. In this work, we identify a new type of structured noise motivated by neutral atom qubits, biased erasure errors, which arises when qubit errors are dominated by detectable leakage from only one of the computational states of the qubit. We study the performance of this model using gate-level simulations of the XZZX surface code. Using the predicted erasure fraction and bias of metastable $^{171}$Yb qubits, we find a threshold of 8.2\% {for two-qubit gate errors}, which is 1.9 times higher than the threshold for unbiased erasures, and 7.5 times higher than the threshold for depolarizing errors. Surprisingly, the improved threshold is achieved without bias-preserving controlled-not gates, and instead results from the lower noise entropy in this model. We also introduce an XZZX cluster state construction for measurement-based error correction, hybrid-fusion, that is optimized for this noise model. By combining fusion operations and deterministic entangling gates, this construction preserves the intrinsic symmetry of the XZZX code, leading to a higher threshold of 10.3\% and enabling the use of rectangular codes with fewer qubits. We discuss a potential physical implementation using a single plane of atoms and moveable tweezers.

\end{abstract}

\maketitle


Quantum error correction (QEC) is essential to protect fragile quantum states during computation \cite{shor1995scheme,shor1996fault, calderbank1996good}. To achieve scalable quantum computation, the rate at which errors are introduced must be below a threshold error rate that depends on the noise model and error correction approach~\cite{aharonov1997fault,knill1998resilient,gottesman1998theory,preskill1998reliable}. Recently, significant work has focused on identifying or engineering the structure of noise in qubits, which can lead to higher thresholds and reduced overhead if paired with appropriate gate operations and quantum error correction (QEC) architectures. For example, biased Pauli noise models can be engineered in superconducting cat qubits~\cite{darmawan2021practical,guillaud2019repetition,chamberland2022building} and certain neutral atom qubits~\cite{cong2022hardware}. Given the availability of bias-preserving gates~\cite{puri2020bias}, this can lead to significantly improved thresholds and lower overhead for the XZZX surface code, which has special symmetries facilitating decoding this type of noise~\cite{Bonilla2021_xzzx,brown2022conservation}. Another example is qubits where errors can be converted with high probability into erasure errors. This model has been proposed for appropriately engineered qubit encodings and gates in neutral atoms~\cite{Wu2022eras}, trapped ions~\cite{Kang2022_ion_era} and superconducting qubits~\cite{kubica2022erasure,teoh2022dual}, and leads to significantly increased thresholds~\cite{stace2009thresholds,barrett2010fault,whiteside2014upper,Wu2022eras,Kang2022_ion_era,kubica2022erasure}

In this work, we identify a new structured error, \emph{biased erasure errors}, that arises when noise is dominated by erasures from only one computational state of the qubit. This model is physically motivated by metastable $^{171}$Yb qubits, where erasures result from leakage out of the $\ket{1}$ computational state into levels whose population can be continuously monitored using cycling transitions that do not affect the qubit levels~\cite{Wu2022eras,campbell2020}. We refer to this as a ($Z$-)biased erasure model, as detecting transitions outside the computational states reveals that the qubit was previously in $\ket{1}$, and can be represented as a $Z$ error with $50\%$ probability. The biased erasure model has more structure than the conventional erasure model, where observing an erasure yields no information about the prior state of the qubit~\cite{grassl1997codes,stace2009thresholds,barrett2010fault,whiteside2014upper,Wu2022eras,Kang2022_ion_era,kubica2022erasure}. Indeed, the error rate at which the quantum capacity of such a channel becomes positive is twice that of the conventional erasure channel, indicating that error correction threshold for the biased-erasure channel can be much higher than that of the conventional erasure channel~\cite{bennett1997capacities}. 

In this work, we study the performance of error correction against biased erasures at the circuit level in several contexts. We first consider the XZZX surface code with conventional, circuit-based syndrome extraction. We find a threshold of 8.2$\%$ when biased erasures comprise $R_e=0.98$ of {the two-qubit} gate errors, as predicted for $^{171}$Yb under optimal conditions~\cite{Wu2022eras}. This is nearly double the threshold of $4.3\%$ for a conventional erasure model, and approximately 8 times the threshold in a comparable depolarizing error model. {The high threshold, compared to depolarizing noise, highlights the benefit of engineering qubits with this favorable error model.} Remarkably, this high threshold is obtained using only the native gate set of neutral atom qubits: single-qubit gates and controlled-Z (CZ) gates, without bias-preserving controlled-not (CX) gates. We attribute the higher threshold to a lower noise entropy when erasures are biased compared to when they are not.

\begin{table*}[ht]
\centering
\begin{tabular}[c]{p{6cm} p{2.0cm} p{2.0cm} p{2.0cm} p{2.2cm} p{2.2cm}}
\toprule[1pt]
\\
 &  &  &  & Preserves 2D   & Avoids atom \\
Model & $R_e=0$ & $R_e=0.98$ & $R_e=1$ & symmetry? & replacement? \\

\toprule[1pt]
\\
circuit, unbiased erasures  &{$1.1\%$} &{$4.3\%$}& {$5.0\%$} & N & N\\
circuit, biased erasures, native gates &{$1.1\%$} & $8.2\%$& $10.3\%$ & N & N\\
circuit, biased erasures, BCX \qquad &{$1.1\%$} & $9.0\%$ & $12.8\%$ & Y & N\\
{hybrid-fusion}, native gates & $1.0\%$ & $10.3\%$ & $14.7\%^\dag$& Y & Y\\
\\
\bottomrule[1pt]
\end{tabular}
\caption{Summary of thresholds derived in this work for the XZZX surface code under various error models and QEC architectures. Thresholds are obtained using an MWPM decoder, and are reported for several values of $R_e$. The first three rows give thresholds using circuit-based syndrome extraction (Sec.~\ref{sec:cqec}), for unbiased erasures~\cite{Wu2022eras}, and using the biased erasure model of Section \ref{sec:bnchan}. The latter error model is studied with and without bias-preserving CX gates, where the former case corresponds to the native gates (single-qubit gates and CZ) of the neutral atom platform. The last line is the hybrid-fusion error correction scheme introduced in Section \ref{sec:fbec}, with native gates. The final two columns indicate additional properties discussed in detail in Section \ref{sec:disc}: whether the dominant errors produce pairs of syndromes lying in 2D planes of the decoding graph (reducing qubit overhead in the limit of large bias), and whether mid-circuit atom replacement is necessary to recover from erasure errors.
$^\dag$This numerically simulated threshold increases to $17.7\%$ when using an erasure decoder at large system sizes (see text).
}
\label{table}
\end{table*}

We also introduce a measurement-based QEC architecture, \emph{hybrid-fusion}, that is specifically tailored for neutral atom qubits with biased erasure noise. This approach combines fusion operations with deterministic entangling gates to construct an XZZX cluster state while preserving the symmetry of the XZZX code under biased noise, without requiring bias-preserving CX gates. In fusion-based (FB) error correction, an error correcting code is built by fusing together few-body entangled resource states using measurements of two-qubit Pauli operators $X\otimes X$ and $Z\otimes Z$. This method has been studied in linear optical quantum computing, where non-deterministic heralded fusions are the native entangling operation~\cite{browne2005resource,li2015resource,bartolucci2021fusion,paesani2022high,sahayfusion}, and has the benefit of preserving the symmetry of the XZZX code when the fusion errors are biased~\cite{sahayfusion}. We present a bias-preserving fusion circuit for qubits with biased erasures, and combine this operation with deterministic entangling gates to develop a measurement-based error correction architecture with a high threshold and reduced overhead. With this approach, we find an even higher threshold of $10.3\%$ for $R_e=0.98$. We also discuss other potential advantages of this approach for neutral atoms including robustness against atom loss and relaxed requirements for erasure detection and atom replacement.

This noise model is physically motivated by metastable $^{171}$Yb qubits~\cite{Wu2022eras}. Recent experimental work has demonstrated high-fidelity gates and validated the basic concept of erasure conversion in this platform, by performing mid-circuit detection of erasure errors with a strong bias from one of the two qubit levels~\cite{
ma2023highfidelity}. Erasure-dominated error models may also be engineered in other qubits with prevalent erasure errors such as metastable trapped-ion qubits~\cite{Kang2022_ion_era}, superconducting qubits encoded in the $\ket{g},\ket{f}$ levels of transmons~\cite{kubica2022erasure}, or dual-rail superconducting qubits~\cite{teoh2022dual} (see Ref.~\cite{lu2023highfidelity, chou2023demonstrating} for recent experimental demonstrations). The high thresholds and reduced requirements for bias-preserving gate operations may encourage the development of new qubits or encodings. Finally, this work may stimulate further development of fusion-based QEC architectures for neutral atoms.
 
The main results of our work are summarized in Table~\ref{table}. We introduce the biased erasure error model in section~\ref{sec:bnchan}, and study its behavior using  circuit-based error correction in section~\ref{sec:cqec}. In  section~\ref{sec:fbec}, we introduce the hybrid-fusion architecture and study its performance. We discuss further opportunities for optimization in section~\ref{sec:disc}, and conclude in section~\ref{sec:con}.

\section{biased erasure errors in neutral atoms}
\label{sec:bnchan}

\begin{figure*}[ht]
    \centering
    \includegraphics[width=6.5in]{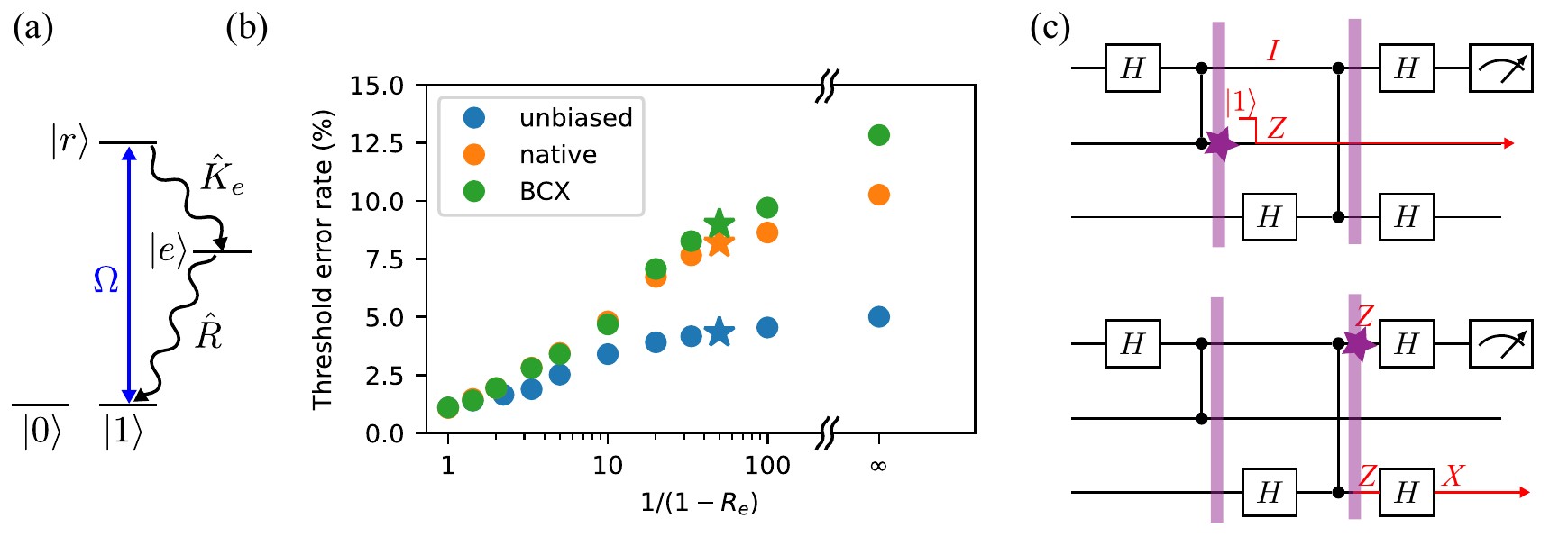}
    \caption{(a) Biased erasures arise when erasures (leakage to a detectable state $\ket{e}$) occur from only one qubit state, $\ket{1}$, such that recovery by replacement or reinitialization in the same state results in at most a $Z$ error. (b) Threshold error rates as a function of $R_e$, under different error models: unbiased erasures (blue), biased erasures with native gates (orange) and biased erasures with bias-preserving CX (BCX) gates (green). The stars denote $R_e = 0.98$. (c) Illustrative circuit measuring the two-qubit stabilizer $ZX$ using native gates. When an erasure is detected during a two-qubit gate (purple star, depicting fluorescence detection of an atom in $\ket{e}$), the affected atom is replaced in $\ket{1}$ and the resulting state is described by an error drawn from $\{I,Z\}^{\otimes 2}$, but the manner in which this error propagates depends on its space-time location in the circuit. If the error occurs during the $Z$ measurement (top), it propagates to the end of the circuit as a Z error. If the error occurs during the $X$ measurement (bottom), it propagates to the end of the circuit as an $X$ error. Knowledge of the error location makes this information available to the decoder, lowering the entropy of the noise.
    }
    \label{fig:171Yb_qubit}
\end{figure*}

To motivate the biased erasure model, we consider how it arises naturally in neutral atom qubits. In this platform, the dominant source of errors are two-qubit gates implemented using the Rydberg blockade~\cite{bluvstein2022,graham2022a,ma2022,schine2022} {(single-qubit and idling errors are comparatively much lower~\cite{ma2023highfidelity})}. The only fundamental effect limiting the fidelity of two-qubit gates is the finite lifetime of the Rydberg state that is populated transiently during the gate~\cite{saffman2010}. In the particular case of $^{171}$Yb~\cite{ma2022,jenkins2022}, encoding the qubit in the nuclear spin sublevels of the metastable $^3P_0$ level has the property that the majority of the decay events populate disjoint subspaces whose occupation can be detected efficiently, which converts these errors into erasure errors~\cite{campbell2020,Wu2022eras}. Recent additional work has derived gate protocols to convert other errors such as quasi-static laser noise and Doppler shifts into erasure errors through a similar mechanism~\cite{jandura2022, fromonteil2022}.

In this work, we consider an additional property of the physical error model of metastable $^{171}$Yb qubits, which is that excitation to the Rydberg state $\ket{r}$ only occurs from the qubit state $\ket{1}$, and never from $\ket{0}$ [Fig.~\ref{fig:171Yb_qubit}(a)] \cite{cong2022hardware}. To illustrate the behavior of this model, consider a hypothetical single qubit operation involving excitation from $\ket{1}$ to $\ket{r}$, where the only possible error is a decay from $\ket{r}$ to a detectable, disjoint state $\ket{e}$. In the absence of an error, the qubit is coherently de-excited back to $\ket{1}$ at the end of the gate. This results in a quantum channel with Kraus operators:
\begin{align}
K_0 &= \ket{0}\bra{0} + \sqrt{1-2p_e}\ket{1}\bra{1} + \ket{e}\bra{e} \\
K_e &= \sqrt{2p_e} \ket{e}\bra{1}
\end{align}
The probability of an error, averaged over both computational states, is $p_e$.

Upon detection of an atom in $\ket{e}$, the qubit is re-initialized into $\ket{1}$, or replaced by a new qubit in $\ket{1}$, described by the recovery operator $\hat{R} = \ket{1}\bra{e}$. The combined channel can be expressed by the Kraus operators:
\begin{align}
\label{eq:w0kraus}
W_0 &= \ket{0}\bra{0} + \sqrt{1-2p_e} \ket{1}\bra{1} \\
W_e &= \hat{R} K_e = \sqrt{2 p_e} \ket{1}\bra{1}
\end{align}

We obtain an effective Pauli channel using the identity $\ket{1}\bra{1} = (I - Z)/2$ and the Pauli twirl approximation (PTA), which may be achieved in practice by inserting random single-qubit Pauli gates after atom replacement~\cite{knill2005quantum,viola2005random,wallman2016noise}. The portion of the channel describing the erasure error is:
\begin{align}
W_e \rho W_e^\dag = \frac{p_e}{2} \left(I \rho I + Z \rho Z\right)
\end{align}
Since the resulting state has at most a $Z$ error, we refer to this as a \emph{biased erasure} error model.

To model a two-qubit CZ gate, we incorporate two additional considerations. First, the leakage of one atom can result in a dephasing error on the other atom as well~\cite{SM}. Therefore, a two-qubit gate with an average erasure probability $p_e$ is modeled by drawing an operator from the set $\{I,Z\}^{\otimes 2}$ with uniform probability $p_e/4$. Second, there is also a finite rate of non-erasure errors such as decays from $\ket{r}$ back to the qubit subspace. We model these as depolarizing errors with total rate $p_p$ by drawing an operator from the set $\{I,X,Y,Z\}^{\otimes 2}\backslash \{I\otimes I\}$ with uniform probability $p_p/15$. The relative probability of these errors is given by the branching ratio $R = p_e/(p_e+p_p)$, which depends on the underlying physics of the qubit. For metastable $^{171}$Yb, we have predicted $R = 0.98$~\cite{Wu2022eras}.

In this work, we only consider errors during two-qubit gates, which are by far the dominant errors for neutral atoms. A discussion of the role of measurement, single-qubit gate and idling errors on the unbiased erasure model can be found in Ref.~\cite{Wu2022eras}.

Lastly, we note that the no-jump error also contributes a $Z$-biased Pauli error with probability $A p_e^2$, because of the asymmetry in the erasure probability from the two qubit states~\cite{cong2022hardware,Wu2022eras}. Continuing the previous example of a single-qubit gate, the evolution under the Kraus operator from Eq.~\eqref{eq:w0kraus} is:
\begin{equation}
\label{eq:w0ch}
W_0 \rho W_0^\dag \approx \left(1-p_e - \frac{p_e^2}{4}\right)I \rho I + \frac{p_e^2}{4} Z \rho Z \\
\end{equation}
Here, we have applied the PTA and taken the limit $p_e \ll 1$.

We find a Pauli error rate of $Ap_e^2$ with $A=1/4$ in Eq.~\eqref{eq:w0ch}, but for the two-qubit gate we estimate $A \approx 1/12$~\cite{SM}. We incorporate this by increasing the Pauli error probability to $p'_p = p_p + A p_e^2$. In the resulting model, erasures constitute a fraction $R_e$ of all errors, with $R_e$ given by:
\begin{equation}
R_e = \frac{p_e}{p'_p + p_e} = \frac{R}{1+A R^2 (p_p + p_e)}
\end{equation}
To present a more generalizable model of biased erasures, we consider the error model to be defined by the independent parameters $R_e$ and $p = p'_p + p_e$. Far below the threshold ($p \ll 1$), the behavior of $^{171}$Yb can be estimated by setting $R_e = R = 0.98$. However, the performance near the threshold will be slightly different, since $R_e < R$. We note that $R_e=1$ is not physically attainable for any value of $R$, but may be achievable in other physical models of biased erasure if both computational states leak with equal rates, but to disjoint final states such that the $Z$ information is preserved.

\section{Circuit-based QEC with biased erasures}
\label{sec:cqec}

We quantify the advantage of the biased erasure model using circuit-level simulations of the XZZX code. The simulations use square codes with distance $d$ up to 13, implemented as $d$ rounds of noisy stabilizer measurements followed by a final, noiseless stabilizer measurement~\cite{SM}.
The error syndromes are decoded using a minimum-weight perfect matching (MWPM) decoder, adjusting edge weights in each shot to incorporate the location of the erasure errors. The stabilizer simulations and construction of the decoding graph are implemented with Stim \cite{Gidney2021_stim}, while the decoding is implemented with PyMatching  \cite{Higgott2022_pymatching}. Except where noted, the simulations do not consider bias-preserving controlled-not (CX) gates. Therefore, CX gates are implemented using controlled-Z (CZ) gates, conjugated by Hadamard (H) gates, which convert $Z$ errors on the target qubit into $X$ errors.

In Fig.~\ref{fig:171Yb_qubit}(b), we show the threshold error rate $p_{th}$ as a function of $R_e$. 
For comparison, we show three cases: unbiased erasures, biased erasures using only the native gates of the Rydberg platform, and biased erasures incorporating hypothetical bias-preserving CX gates. 

For large values of $R_e$,  biased erasures result in  significantly higher thresholds than unbiased erasures, even in the absence of bias preserving gates. For example, at the value $R_e = 0.98$ projected for metastable $^{171}$Yb qubits, the threshold is $8.2\%$, nearly twice the value with unbiased erasures ($p_{th} = 4.3\%$) and nearly eight times the threshold with depolarizing noise ($R_e=0$, $p_{th} = 1.1\%$). The latter two thresholds are slightly higher than those reported in Ref. \cite{Wu2022eras}, because we use the slightly more accurate MWPM decoder, instead of a weighted Union Find decoder.

Previous works using the XZZX surface code to correct biased Pauli noise leverage its particular symmetry which guarantees that the pairs of error syndromes created by $Z$ errors on the data and ancilla qubits lie in disconnected 2D planes~\cite{Bonilla2021_xzzx,brown2022conservation}. This advantage  vanishes in the absence of bias-preserving gates (see, for instance, Ref.\,\cite{darmawan2021practical}). The fact that we observe high thresholds with only the native gates, and relatively little additional improvement from incorporating bias-preserving CX gates for $R_e < 1$ [Fig.~\ref{fig:171Yb_qubit}(b)], suggests that other mechanisms are responsible. This is reinforced by a separate calculation showing that the CSS and XZZX surface codes give almost the same threshold for biased erasures with native gates at $R_e = 0.98$.

The high threshold with native gates arises from two mechanisms. First, the biased erasure model has a lower error probability than the unbiased erasure model: after returning to the qubit space, the probability of an error in the biased erasure model is $3/4$, compared to $15/16$ for the unbiased model. Second, even though $Z$ errors can be converted into $X$ errors in the absence of bias-preserving gates, detecting erasures after every gate allows this evolution to be tracked, lowering the entropy of noise [Fig.~\ref{fig:171Yb_qubit}(c)] and reducing the impact of bias-preserving gates on the threshold.

Previous works on biased Pauli errors have also proposed using a thin rectangular XZZX code with a smaller distance for the low-rate error~\cite{Bonilla2021_xzzx,darmawan2021practical}, which provides an additional reduction in qubit overhead. This is not possible in the biased erasure model, without bias-preserving CX gates, because the dominant $Z$ errors get converted to $X$ errors. However, we will see in the next section that this can be overcome using an alternate approach based on fusions.

For estimating the performance of biased erasure models in other qubit platforms that may have varying degrees of bias, and for including potential bias-degrading effects in $^{171}$Yb, we have parameterized a finite bias version and computed thresholds as a function of bias in the supplementary information \cite{SM}. We find that the rate of erasures from the low-probability state (here, $\ket{0}$) must be $\sim 100$ times less than the high-probability state to take full advantage of the bias (similar to case of biased Pauli noise~\cite{Bonilla2021_xzzx,darmawan2021practical}). {While these simulations do not include errors in single-qubit gates, measurements, or idling qubits, we believe that these will not have a significant effect because these operations are comparatively higher fidelity for neutral atom qubits~\cite{ma2023highfidelity}.}

\section{Hybrid-fusion QEC}
\label{sec:fbec}

\begin{figure*}[t]
\includegraphics[width=\textwidth]{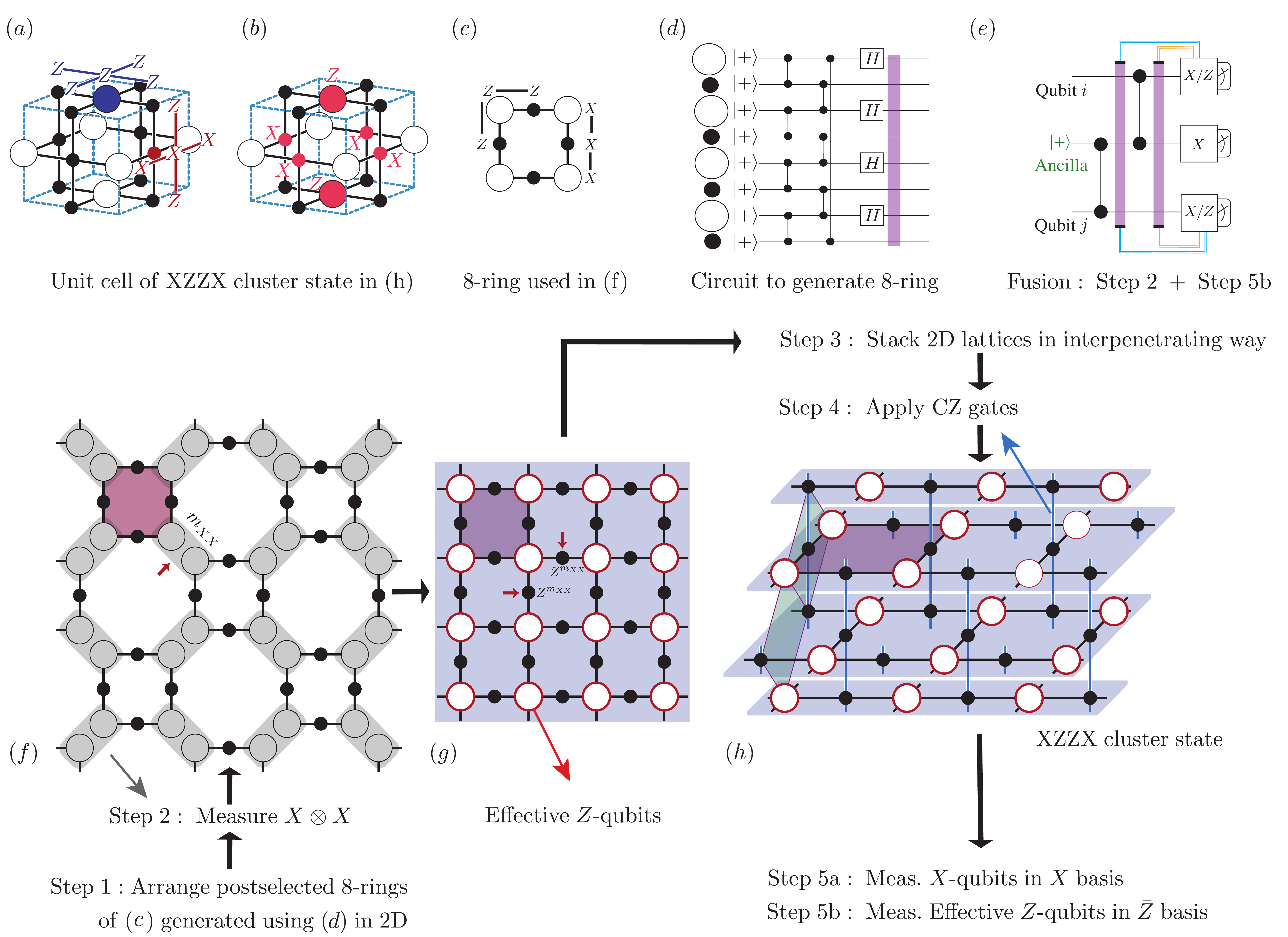}
\caption{(a) A unit cell of the XZZX cluster state, with two
examples of stabilizers centered on $X$ ($\newmoon$)- and $Z$ ($\bigcirc$)- type qubits as described in the main text. (b) The cell stabilizer obtained by multiplying the stabilizers centered at all faces of a unit cell.
(c) The 8-ring resource states used to build the cluster state. (d) Circuit to generate the resource state with $^{171}$Yb atoms. The state is postselected on the absence of detected erasure errors. We have drawn only a single erasure detection step at the end of the circuit to reflect that the precise space-time location of the errors is not needed. In practice, a fluorescence detection is performed after every gate. (e) Circuit for adaptive, bias-preserving fusion measurements with $^{171}$Yb atoms. Erasure detection (via fluorescence, purple lines) is performed following each gate. If any erasures occur, the measurement basis of the fusion qubits is changed from $X$ to $Z$, ensuring that the value of $Z \otimes Z$ is preserved at the expense of $X \otimes X$.  (f-h) Extended protocol for conceptual understanding of how the cluster state error correction is realized. For reference, in (h) we highlight in green a planar array of qubits encoded in XZZX surface code that is being propagated in time (left to right), generating the XZZX cluster state. Note that the entire cluster state shown does not need to be built at once, and can be realized using a small number of such planar arrays of qubits which are reused over time (see~\cite{SM}). }
\label{fig:combined}
\end{figure*}

 Measurement-based error correction (MBEC) is an alternative approach to error correction based on performing local measurements on a many-body 3D entangled state called a fault-tolerant cluster state~\cite{briegel2001persistent,raussendorf2001one,raussendorf2002one,raussendorf2003measurement,briegel2009measurement}. In the standard approach to realize a fault-tolerant cluster state,
called {\it foliation}, one dimension of the cluster state effectively simulates time along which a planar encoded state is propagated via teleportation~\cite{raussendorf2006fault,raussendorf2007fault,bolt2016foliated,brown2020universal}. Indeed, the commonly used Raussendorf-Harrington-Goyal (RHG) or Raussendorf-Bravyi-Harrington cluster state ~\cite{raussendorf2006fault,raussendorf2007fault,raussendorf2005long} teleports the standard CSS surface code. Similarly, the recently introduced XZZX cluster state teleports the XZZX surface code~\cite{claes2022tailored}. {Given planar arrays with a finite number of qubits, a computation of length scaling exponentially with the size of the array can be performed by using as few as two of these arrays at a time~\cite{raussendorf2007fault}, and teleporting the logical state back and forth between them.} After a slice is measured, it is re-initialized into a cluster state.

Fusion-based error correction (FBEC) is a particular approach for MBEC in which the cluster state is grown by fusing together few-body entangled resource states. Fusions are entangling operations carried out by performing destructive two-qubit measurements of $X\otimes X$ and $Z\otimes Z$. FBEC has been widely studied for linear optical quantum computing  because fusions are the native entangling operations in that platform~\cite{browne2005resource,li2015resource,bartolucci2021fusion,paesani2022high,sahayfusion}. {Recently, a fusion-based construction of the XZZX code was proposed that can maintain the symmetry of that code under biased noise if the fusion errors are biased~\cite{sahayfusion}.}

In this section, we introduce a \emph{hybrid-fusion} construction of the XZZX cluster state with biased erasures that combines fusion operations and deterministic entangling gates, which is optimized  for the error model and capabilities of metastable $^{171}$Yb qubits. After reviewing the XZZX cluster state, we introduce an 8-qubit resource state (hereafter, \emph{8-ring}) and present its construction. Next, we design a {bias-preserving fusion circuit that ensures that biased erasures in the physical gates only affect the $X\otimes X$ measurements, preserving $Z\otimes Z$}. Finally, we present the hybrid cluster state construction protocol that uses both fusions and direct CZ gates {to entangle a collection} of 8-ring resource states. The operations and resource states are designed to minimize the number of CZ gates while ensuring that biased erasures at any step maintain the two-dimensional symmetry of the XZZX code. This simplifies the decoding problem, enabling higher thresholds and lower overhead~\cite{sahayfusion,brown2022conservation}, which we demonstrate using circuit-level simulations.

\subsection{The XZZX cluster state}

The XZZX cluster state~\cite{claes2022tailored} is a stabilizer state defined on a graph with an {\it $X$-type} or {\it $Z$-type} qubit at each vertex, represented by $\newmoon$ and $\bigcirc$ respectively in Fig.~\ref{fig:combined}(a). There is a stabilizer centered at each vertex. The stabilizer centered at a $X$-type qubit is the product of the Pauli $X$ operator of that qubit, the Pauli $Z$ operators of all adjacent $X$-type qubits and the Pauli $X$ operators of all the adjacent $Z$-type qubits. The stabilizer centered at a $Z$-type qubit is the product of the Pauli $Z$ operator of that qubit and the Pauli $Z$ operators of all adjacent $X$-type qubits. There is no edge between two $Z$-type qubits in the states considered here. Multiplying the stabilizers centered on the faces of a unit cell gives the six-body cell stabilizer, which is a product of Pauli $X$ operators on the $X$-type qubits and Pauli $Z$ operators on the $Z$-type qubits on the faces of the cell, as shown in Fig.~\ref{fig:combined}(b). Measuring all $Z$-type qubits in the $Z$ basis and all $X$-type qubits in the $X$ basis teleports the XZZX surface code through this cluster state. 

In the XZZX cluster state, a $Z$ error on an $X$-type qubit or an $X$ error on a $Z$-type qubit causes its neighboring cell stabilizers to flip~\cite{claes2022tailored}. To simultaneously perform error correction, the value of each cell stabilizer is constructed by adding the measurements outcomes of qubits around the faces of the unit cells. Importantly, $Z$ errors on $X$-type qubits only cause pairs of defects, or error syndromes, that are confined to lie in disconnected 2D layers, leading to more accurate decoding of errors and higher thresholds~\cite{Bonilla2021_xzzx,claes2022tailored,brown2022conservation}. This feature reflects the symmetry that arises within the stabilizer group of the XZZX code under $Z$ errors. To take advantage of this feature (\emph{i.e.}, by using a rectangular code), it is necessary to ensure that the dominant physical noise during cluster state preparation preserves this symmetry, by only introducing $Z$ errors on $X$-type qubits.

This property is satisfied when using CZ gates with biased erasures to directly entangle $X$-type qubits. However, without a bias-preserving CX gate, it is not possible to directly entangle $X$- and $Z$-type qubits without converting the dominant $Z$-biased erasures to $X$-biased erasures and thus introducing unwanted $X$ errors on the $Z$-type qubits. To overcome this challenge, our approach is to isolate the generation of entanglement between $X$- and $Z$-type qubits in the creation of 8-ring resource states that are post-selected on the absence of erasures. These are then joined together into layers using adaptive fusion measurements which preserves the noise bias so that dominant $Z$-biased erasures do not introduce errors on $Z$-type qubits. Finally, the layers are joined to form the 3D cluster state using only CZ gates on $X$-type qubits.

\subsection{The resource state }
The cluster state is assembled out of a collection of 8-ring resource states. The 8-ring state is defined by the graph  in Fig.~\ref{fig:combined}(c), and can be prepared using the circuit in Fig.~\ref{fig:combined}(d). This circuit involves 8 CZ gates between neighboring $X$- and $Z$-type qubits on a ring. Biased erasures can result in unwanted $X$ errors on $Z$-type qubits. However, postselecting completed rings on the absence of erasures allows these errors to be removed, while at the same time increasing the overall fidelity of the resource state. Using the notation of Section \ref{sec:bnchan}, in the limit where $R_e\approx 1$ and $p_e\ll 1$, the success probability is $1-8 p_e$ and the error probability of successful resource states is $\approx 8 p'_p$. Many copies of this resource state can be prepared in parallel, and the successful ones can be moved into the positions for cluster state construction, described next, using movable optical tweezers~\cite{beugnon2007,bluvstein2022}.

\subsection{Adaptive, bias-preserving fusion measurements}\label{ssec:fusions}
A fusion measurement is a destructive two-qubit measurement of $X\otimes X$ and $Z\otimes Z$. Figure~\ref{fig:combined}(e) presents an adaptive fusion measurement circuit with the property that biased erasures during two-qubit gates only cause an erasure of the $X\otimes X$ measurement outcome. In the ideal evolution without errors, the ancilla qubit measures $Z_i \otimes Z_j$ using CZ gates, followed by the single-qubit measurements $X_i\otimes I_j$ and $I_i\otimes X_j$, from which $X_i \otimes X_j$ can be computed~\cite{SM}. 

In order to concentrate dominant errors into the $X_i\otimes X_j$ measurements and preserve the $Z_i \otimes Z_j$ information, we check for erasure errors in the two CZ gates, as shown in Fig.~\ref{fig:combined}(e), and adapt the subsequent operations based on the location of these errors. In particular, when an erasure is detected, the protocol is aborted and each fusion qubit is measured independently in the $Z$ basis with the measurement outcomes $m_{i}$ (= 0 or 1) and $m_{j}$ (= 0 or 1). The overall evolution is as if the atoms $i,j$ were fused, with the $Z_i\otimes Z_j$ measurement outcome $=m_{i}\oplus m_{j}$, but the $X_i\otimes X_j$ measurement outcome is erased. We note that the $Z$ measurement only needs to be performed on the fusion qubit that is not erased, because observing the erasure is equivalent to measuring the erased atom in $\ket{1}$. Therefore, the erased qubits also do not need to be replaced, unlike the approach in Section \ref{sec:cqec} and Ref.~\cite{Wu2022eras}.

\subsection{Constructing the XZZX cluster state} \label{sec:constXZZXCS}
We now present a technique to construct the full XZZX cluster state using 8-rings, fusion measurements, and deterministic entangling gates. We first present the most intuitive version of the protocol (Fig.~\ref{fig:combined}f-h), and then a contracted version (Fig.~\ref{fig:short}).

\subsubsection{Intuitive protocol}
In {\it step 1}, copies of postselected 8-ring resource states [Fig.~\ref{fig:combined}(c,d)] are arranged in a plane as shown in Fig.~\ref{fig:combined}(f). In {\it step 2}, an $X\otimes X$ measurement is performed on pairs of $Z$-type qubits at the neighboring corners. This measurement joins the pair of measured qubits into a single effective $Z$-type qubit with logical $Z$ operator $\bar Z=Z\otimes Z$, giving a single 2D layer of the XZZX cluster state shown in Fig.~\ref{fig:combined}(g)~\cite{bartolucci2021fusion,sahayfusion}. To ensure that the post-measurement state is the stabilizer state defined by the graph in Fig.~\ref{fig:combined}(g), a Pauli $Z$ correction is applied to the two $X$-type qubits adjacent to one of the measured $Z$-type qubits conditional on the outcome of the $X\otimes X$ measurement~\cite{sahayfusion}. In practice, this correction is tracked in software.

In {\it step 3}, copies of such 2D lattices are stacked on top of each other in a staggered manner such that the $X$-type qubits in one layer align with those in the next, while the $Z$-type qubits in one layer lie on top of a face in the next layer [Fig.~\ref{fig:combined}(h)]. In {\it step 4}, a CZ gate is applied between each $X$-type qubit in layer $k$ and another $X$-type qubit at the same location in layer $k+1$ as shown in Fig.~\ref{fig:combined}(h). This gives the entire 3D XZZX cluster state. Importantly, the CZ gates commute with each other and may be applied in any order. Here we follow a specific order: for each $X$-type qubit connected to unit cells to its left and right, the CZ gate with the $X$-type qubit in the layer above it is performed before the CZ gate with the layer below. If an erasure is detected in the first CZ gate, the second CZ gate is omitted to avoid introducing additional errors.

Now that we have the cluster state, we measure each qubit to teleport the XZZX surface code through the cluster state and to reconstruct the cell stabilizers of Fig.~\ref{fig:combined}(b) for error correction. This is divided into two sub-steps. In {\it step 5a}, we measure each $X$-type qubit in $X$ basis. In {\it step 5b}, we measure each effective $Z$-type qubits in the effective $\bar Z$-basis, by measuring $Z\otimes Z$ on the physical $Z$-type qubits composing the effective qubit. This outcome is not affected by biased erasures, since the fusion circuit ensures that the $Z\otimes Z$ result is preserved.

\subsubsection{Contracted protocol}
\begin{figure}
\includegraphics[width=\columnwidth]{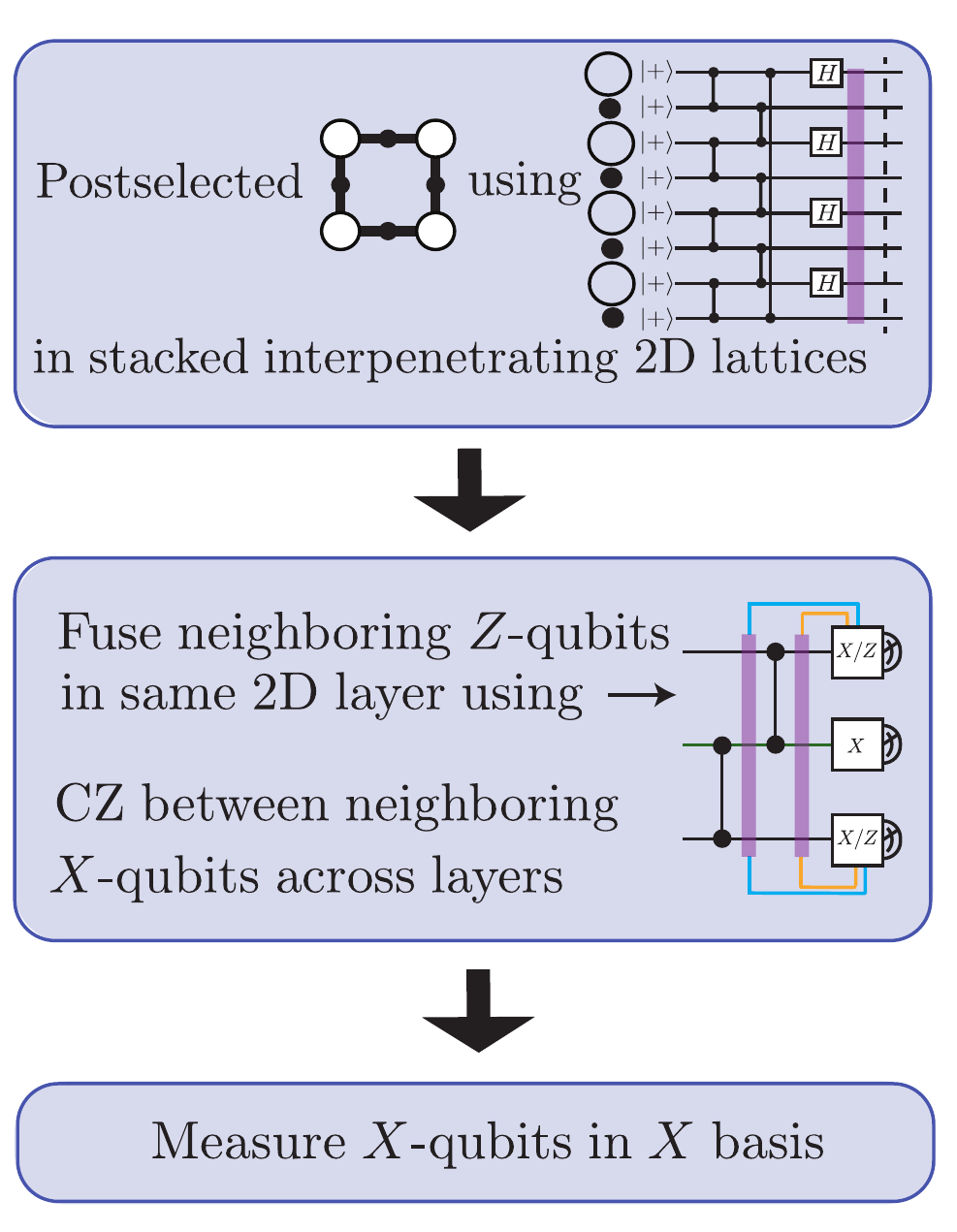}
\caption{Proposed shortened protocol for high-threshold hybrid-fusion QEC with the XZZX cluster state using fusions and CZ gates. }
\label{fig:short}
\end{figure}

We now observe that these operations can be regrouped to shorten the protocol. First, step 5b commutes with steps 5a, 4, 3, and 2, and can therefore be performed simultaneously with step 2. Steps 5b and 2 together constitute a fusion measurement, which removes $Z$-type qubits from the cluster state entirely. This fusion measurement is implemented using the circuit in Fig.~\ref{fig:combined}(e). Furthermore, the staggered layer stacking and CZ gates in steps 3 and 4, respectively, can also be performed concurrent with or before the fusion measurements, as they act on a different subset of the qubits. Thus, the sequence of operations in the shortened protocol, summarized in Fig.~\ref{fig:short}, begins by preparing several copies of post-selected resource states and moving them in position to form layers stacked in a staggered manner (steps 1,3), which is followed by fusions (steps 2, 5b) and CZ gates between layers (step 4), and finally measurement of $X$ qubits in the $X$ basis (step 5a). {Note that, in the case of a biased erasure during a fusion, we do not obtain the $X\otimes X$ measurement information, which means we cannot determine the Pauli correction on the two neighboring $X$-type qubits as discussed in section \ref{ssec:fusions}. This is effectively a random Pauli $Z\otimes Z$ error on these qubits. This correlated two-qubit error does not reduce the distance, as it is not oriented along a logical operator. }

{We also point out that accumulation of coherent errors, such as from the no-jump evolution, will be suppressed because every qubit is measured frequently (at most after 4 CZ gates). As with the circuit-based approach, additional suppression can be realized by twirling, inserting random Pauli gates before the CZ gates in resource state generation, fusion circuit and in interlayer entangling steps. }

 \begin{figure}
    \centering
\includegraphics[width=\columnwidth]{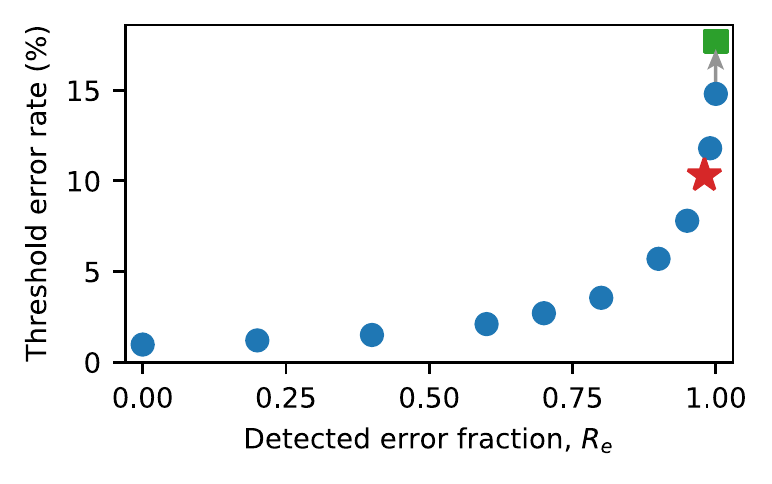}
    \caption{Threshold error rates for the hybrid-fusion architecture as a function of erasure fraction, $R_e$. The star denotes $R_e=98\%$, while the green square marks the percolation threshold corresponding to the decoding graph for $R_e = 1$ (see text).
    }
    \label{fig:threshold_R_clusten}
\end{figure}

\subsection{Threshold results}

We evaluate the performance of our hybrid-fusion architecture under the biased erasure noise model by estimating the thresholds for different $R_e$ (Fig.~\ref{fig:threshold_R_clusten}). Our simulations account for noise in the CZ gates used in the resource state generation circuit, the fusion circuit, and the direct entanglement in step 4. Errors are decoded using the MWPM decoder and a $d\times d\times d$ cluster state, with $d$ up to $13$, which is equivalent to teleporting a $d\times d$ planar XZZX code while performing $d$ rounds of stabilizer measurements in the circuit-based approach.

For $R_e=0.98$, we obtain a threshold of $10.3\%$, higher than that achieved with the circuit-based approach with or without bias-preserving CX gates. In the other extreme of $R_e=0$ when all source of noise is depolarizing Pauli noise, we obtain a threshold of $1\%$, similar to the threshold with circuit-based error correction at the same value of $R_e$.

In the limit $R_e=1$, we achieve a threshold of $14.7\%$. This threshold can also be determined from the bond-percolation threshold of the union-jack lattice and is expected to be $\sim 17.7\%$ \cite{SM}. The observed threshold in Fig.~\ref{fig:threshold_R_clusten} is smaller than this because of finite size effects that are more prominent in the extreme case of $R_e=1$ when the decoding graph is 2D. To confirm, we use a fast erasure decoder ~\cite{delfosse2020linear} to simulate extremely large lattices ($d\times d\times 5$  up to $d=61$), and recover the threshold predicted by percolation theory in this limit. Since the erasure decoder and MWPM decoder achieve the same accuracy for erasure errors, we believe that we should achieve the same thresholds for both decoders at $R_e=1$, but simulating such large lattices with MWPM is computationally prohibitive.

\subsection{Comparison to other approaches to generating a cluster state}
The benefit of the hybrid-fusion approach can be understood by comparison to alternative cluster state constructions. Compared to directly entangling all the qubits in the cluster state, our hybrid-fusion approach results in the same number of biased erasure errors on the final cluster state, but ensures that the error syndromes lie in disconnected 2D layers. However, the number of CZ gates, and thus the rough Pauli error rate on the final cluster state, is increased by a factor of $4/3$. {Therefore, our approach should outperform direct entanglement except when $R_e$ is very small} {and Pauli errors are more important to correct than biased erasures}.

On the other hand, we can compare to an all-fusion strategy like the 6-ring construction of Ref.~\cite{sahayfusion}. Our strategy again has the same number of biased erasure errors on the final cluster state, but uses $2/3$ the number of CZ gates. {Therefore, our approach outperforms the all-fusion construction except when $R_e \gtrsim 0.99$, where the higher percolation threshold of the decoding graph in the all-fusion approach gives a slight advantage.}

\section{Discussion}
\label{sec:disc}

We now make several comparisons between the presented circuit-based and hybrid-fusion approaches. Our hybrid-fusion protocol can be viewed as implementing QEC on a state encoded in a planar XZZX code as it is being teleported to another planar array of qubits. This is an alternative to using repeated rounds of quantum nondemolition stabilizer measurements as in conventional circuit-based error correction, but still allows transversal one- and two-qubit logical gate operations available for the planar surface code. In the supplementary material, we explicitly show how our hybrid-fusion protocol can be implemented with just a few planar arrays of qubits, starting and stopping with a 2D encoded XZZX surface code~\cite{SM}. We note that neutral atoms are ideally suited to the high degree of connectivity required to implement hybrid-fusion: dynamic rearrangement of qubits is already used to postselect filled tweezer sites~\cite{endres2016,barredo2016}, and coherent qubit transport has also been demonstrated~\cite{beugnon2007,bluvstein2022}.

The hybrid-fusion approach has other advantages beyond the higher threshold, which we outline here but do not quantitatively analyze. As discussed, our construction preserves the system symmetry of the XZZX code under biased noise by ensuring that the high-probability $Z$ erasures run along layers in Fig.~\ref{fig:combined}(h). Thus, the same logical error rate can be achieved with fewer layers when $R_e$ is large and the erasure are highly biased. This amounts to using a thin, rectangular XZZX surface code, which becomes a repetition code when $R_e=1$ and the erasures are infinitely biased \cite{Bonilla2021_xzzx}. This allows for reduced overhead compared to the circuit-based approach, which requires a square XZZX code in the absence of bias-preserving gates. This property is summarized for each QEC architecture in Table \ref{table}.

Second, the teleportation process converts lost atoms into Pauli errors, ensuring a finite threshold against loss errors or undetected erasures without additional leakage reduction units~\cite{knill2005scalable,aliferis2005fault,fowler2013coping,ghosh2015leakage,suchara2015leakage}. We leave an analysis of the threshold for a given loss rate to future work.

Finally, hybrid-fusion QEC relaxes the requirements for erasure detection and subsequent atom replacement. In the circuit-approach proposed in Ref. \cite{Wu2022eras} and considered in Section \ref{sec:cqec}, the space-time location of each erasure error in the circuit must be resolved, and the affected qubits must be replaced or re-initialized as the computation proceeds. In the hybrid-fusion approach, this requirement is relaxed, in a way that is slightly different for the three steps involving two-qubit gates. For resource state preparation (step 1 of the protocol in Sec.~\ref{sec:constXZZXCS}), it is only necessary to determine if an erasure occurred at some point during the 8-ring preparation, and if it did, the entire state is discarded. Therefore, there is no need to replace affected qubits, and the necessary spatio-temporal resolution of the erasure detection is significantly coarser. During the fusion operations (steps 2, 5b), erasures must be detected immediately as the measurement basis is conditioned on this outcome, but the affected atoms do not need to be replaced. Finally, in the layer-joining CZ gates (step 4), atom replacement is not necessary, as the atoms are immediately measured. In summary, we find that conditionally replacing atoms at precise space-time locations is never required (as indicated in Table \ref{table}), and in some cases erasures can be detected with coarser resolution. These may give rise to considerable experimental simplifications.

\section{Conclusion}
\label{sec:con}
To summarize, we have introduced a new noise model, biased erasure, that is physically motivated by metastable $^{171}$Yb qubits but may also be engineered in other qubit platforms. We have studied two realistic QEC architectures under this noise model. The first is a circuit-based approach, where the improvement with the biased erasure model arises from the reduced entropy of this noise model, enabling more effective decoding. We obtain a threshold of 8.2\% for the predicted metastable $^{171}$Yb erasure fraction. The second is a hybrid-fusion approach with a systematic code construction that gives rise to system symmetries under the biased erasure model. In this approach, we obtain a threshold of 10.3\% for the metastable $^{171}$Yb noise model. Compared to circuit-based syndrome extraction, this approach has the additional benefits of potentially enabling rectangular surface codes with lower overhead (in the limit of a large erasure fraction), robustness against atom loss, and simplified requirements for detecting and handling erasure errors in real-time.

{While in this work we have focused on thresholds for quantum memory, we can apply Clifford gates using standard techniques such as braiding~\cite{raussendorf2006fault,raussendorf2007fault}, lattice surgery~\cite{horsman2012surface,litinski2019game}, or other code deformations~\cite{PhysRevX.7.021029}. These can be straightforwardly adapted to our hybrid-fusion approach by modifying the large-scale shape of our cluster state, but using the same underlying resource states and without affecting the thresholds~\cite{herr2018lattice,brown2020universal,bartolucci2021fusion}. In addition, we can consider entangling logical surface-code qubits via transversal CNOT gates as proposed in~\cite{dennis2002topological}, thanks to the movability of neutral atom qubits~\cite{beugnon2007,bluvstein2022}; this may allow for significant overhead reduction compared to lattice surgery. To apply non-Clifford gates, we can inject noisy magic states and perform magic state distillation~\cite{bravyi2005universal}.  We note that the erasure errors during magic state injection may be removed by post-selection, so that injected magic states are only susceptible to low-probability Pauli errors. This will aid in reducing the overhead cost of magic state distillation.  }

There are several opportunities for further simplification and optimization of the hybrid-fusion approach for scalable fault-tolerant quantum computing. First, the threshold for hybrid-fusion QEC may be further improved by post-selecting on larger resource states~\cite{paesani2022high}. This will decrease the rate of successfully generating the resource states, but it is still an experimentally viable route with  deterministic, high-fidelity gates. For example, given a gate with 99.9$\%$ CZ fidelity, resource state chunks involving 100 CZ gates could be post-selected with 90\% success probability. {Second, we can explore error correction beyond the planar surface code, as} the hybrid-fusion approach naturally allows conversion of the planar surface code to a 3D surface code with transversal non-Clifford gates{~\cite{bombin20182d,vasmer2019three,brown2020fault,scruby2022numerical,cai2022looped}}, which may lead to significant overhead reduction compared to alternative protocols for non-Clifford gates that rely on magic-state distillation.

{Finally, our hybrid-fusion-based construction provided a means to preserve the symmetries of the XZZX code without bias-preserving CX gates by using post-selection and short-depth circuits. This approach also provides a path for high-threshold QEC with biased-Pauli-noise qubits without bias-preserving CX gates, which is an open challenge so far.}

\section{Acknowledgements}
We acknowledge Steve Flammia for pointing out the importance of lowered noise entropy. We are also grateful to Shimon Kolkowitz, Steve Girvin, Shilin Huang, Shuo Ma, and Yue Wu for helpful discussions. 
This work was supported by the National Science Foundation (QLCI grant OMA-2120757). JDT acknowledges additional support from ARO PECASE (W911NF-18-10215), ONR (N00014-20-1-2426), DARPA ONISQ (W911NF-20-10021) and the Sloan Foundation.

\bibliography{references}

\setcounter{equation}{0}
\setcounter{figure}{0}
\setcounter{table}{0}
\setcounter{section}{0}
\makeatletter

\renewcommand{\theequation}{S\arabic{equation}}
\renewcommand{\theHequation}{S\arabic{equation}}
\renewcommand{\thefigure}{S\arabic{figure}}
\renewcommand{\theHfigure}{S\arabic{figure}}
\renewcommand{\thetable}{S\arabic{table}}
\renewcommand{\theHtable}{S\arabic{table}}
\renewcommand{\thesection}{S\arabic{section}}
\renewcommand{\theHsection}{S\arabic{section}}

\title{Supplementary Material: High threshold codes for neutral atom qubits with biased erasure errors}

\clearpage
\maketitle
\onecolumngrid


\section{Error channel}

\subsection{Two-qubit gate}
Following the description of the single-qubit gate error channel in Section~\ref{sec:bnchan} in the main text, here we consider the more realistic two-qubit gate. Decay of one or both atoms to $\ket{e}$ can occur during the gate, which  {we approximate} by the channel with Kraus operators:
\begin{subequations}
\begin{align}
K_0 = \ket{00}\bra{00} + \sqrt{1-p_1} \left(\ket{10}\bra{10} + \ket{01}\bra{01}\right) &+ \sqrt{1-p_2} \ket{11}\bra{11} + \ket{e}\bra{e} \otimes I + I \otimes \ket{e}\bra{e} + \ket{ee}\bra{ee}\\
K_1 &= \sqrt{p_1} \ket{e0}\bra{10} \\
K_2 &= \sqrt{p_1} \ket{0e}\bra{01} \\
K_3 &= \sqrt{p_2/3} \ket{e1}\bra{11} \\
K_4 &= \sqrt{p_2/3} \ket{1e}\bra{11} \\
K_5 &= \sqrt{p_2/3} \ket{ee}\bra{11}
\end{align}
\end{subequations}

Here, $p_1$ is the probability of decaying to $\ket{e}$ from the states $\ket{01}$ or $\ket{10}$, and $p_2$ is the probability for any decay to $\ket{e}$ from $\ket{11}$. 
 {For typical CZ gates, $p_1 \approx p_2$ because the Rydberg blockade results in only one atom in $\ket{r}$ in either case~\cite{Levine2019_para_cz,Wu2022eras}. The total erasure probability is defined} by averaging over the computational states as $p_e = (2 p_1 + p_2)/4$.

After detecting whether any atoms are in $\ket{e}$, we apply one of the following recovery operators:

\begin{subequations}
\begin{align}
R_0 &= I \\
R_1 &= \ket{1}\bra{e} \otimes I \\
R_2 &= I \otimes \ket{1}\bra{e} \\
R_3 &= \ket{11} \bra{ee}
\end{align}
\end{subequations}

This results in {a channel with Kraus operators}:

\begin{subequations}
\begin{align}
\label{eq:w0}
W_0 = R_0 K_0 &= \ket{00}\bra{00} + \sqrt{1-{p_1}} \left(\ket{10}\bra{10} + \ket{01}\bra{01}\right) + \sqrt{1-p_2} \ket{11}\bra{11} \\
W_1 &= R_1 K_1 = \sqrt{p_1} \ket{10}\bra{10} =  \frac{\sqrt{p_1}}{4} (I-Z) \otimes (I+Z) \\
W_2 &= R_2 K_2 = \sqrt{p_1} \ket{01}\bra{01} =  \frac{\sqrt{p_1}}{4} (I+Z) \otimes (I-Z) \\
W_3 &= R_1 K_3 = \sqrt{p_2/3} \ket{11}\bra{11} =  \sqrt{\frac{p_2}{3}} \frac{1}{4} (I-Z) \otimes (I-Z) \\
W_4 &= R_2 K_4 = W_3 \\
W_5 &= R_3 K_5 = W_3 
\end{align}
\end{subequations}

Applying the PTA as before yields the Pauli channel:

\begin{subequations}
\begin{align}
\label{eq:w0final}
W_0 \rho W_0^\dag &\approx {N} II \rho II + \frac{p_2^2}{64} \left(IZ \rho IZ + ZI \rho ZI \right) + \frac{({2p_1 - p_2})^2}{64} ZZ \rho ZZ \\
W_1 \rho W_1^\dag = W_2 \rho W_2^\dag &\approx \frac{p_1}{16} \left(II \rho II + IZ \rho IZ + ZI \rho ZI + ZZ \rho ZZ\right) \\
W_3 \rho W_3^\dag = W_4 \rho W_4^\dag = W_5 \rho W_5^\dag &\approx \frac{p_2/3}{16} \left(II \rho II + IZ \rho IZ + ZI \rho ZI + ZZ \rho ZZ\right) 
\end{align}
\end{subequations}

{where $N$ is a normalization factor}. The resulting qubit state is the same in all of the channels corresponding to an erasure detection ($i \geq 1$). Therefore, we group them together and consider their sum:

\begin{equation}
\label{eq:finalbias}
\sum_{i \geq 1} W_i \rho W_i^\dag = \frac{p_e}{4} \left(II \rho II + IZ \rho IZ + ZI \rho ZI + ZZ \rho ZZ\right)
\end{equation}
which is the error model described in the main text.

The no-jump error when no erasure is detected is slightly more complicated than in the single qubit case {presented in Sec.~\ref{sec:bnchan}}. The physical interpretation of this error is that the absence of a detected error reveals information about the state, specifically, that it is less likely to be in a state from which it could have jumped to $\ket{e}$.

{We note that the last coefficient in Eq.~\eqref{eq:w0final} would vanish when the two qubits jump independently ($p_2 = 2 p_1$), indicating no correlated errors up to the second order.}
If $p_2 < 2 p_1$ (which is the case in practice, because of the Rydberg blockade), then the extra weight of the $ZZ$ term in Eq.~\eqref{eq:w0final} represents a correlated error, {but we do not consider this aspect further since these errors are already very small below threshold. Using the definition of $p_e$ and Eq.~\eqref{eq:w0final} with $p_1 = p_2$, we arrive at the estimated no-jump Pauli error contribution of $p_e^2/12$ that is referenced in Sec.~\ref{sec:bnchan}.}

\subsection{Erasures with finite bias}
\label{subsec:finite_bias}

\begin{figure}
\centering
\includegraphics[width=\columnwidth]{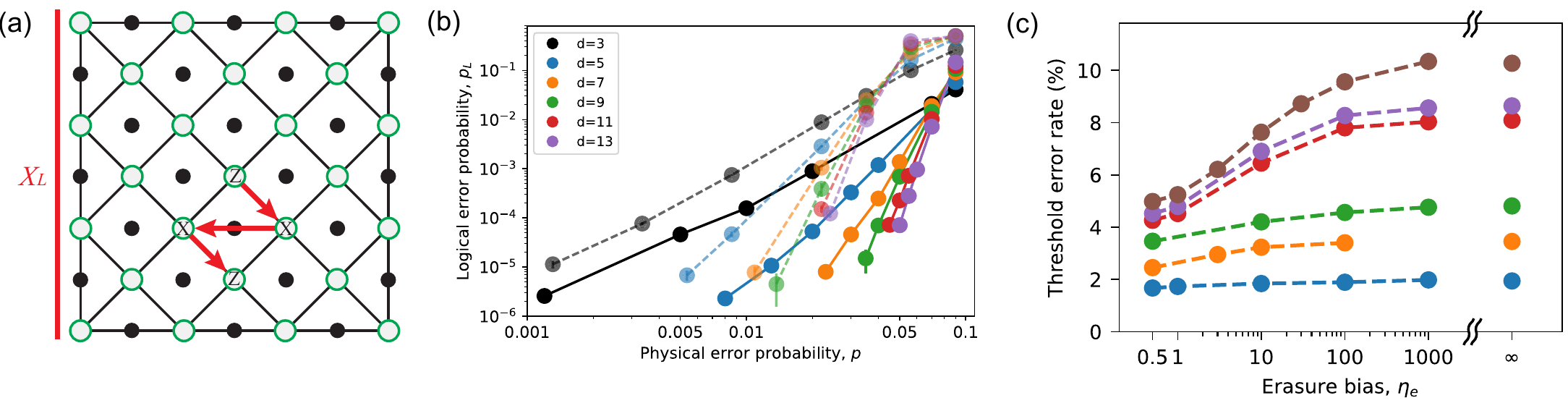}
\caption{(a) A patch of the XZZX surface code studied in this work, with the data and ancilla qubits denoted by the green open circles and black solid circles, respectively. The red arrows denote the sequence of the stabilizer measurement. (b) Scaling of the logical error rate with the physical error rate for code distances up to $d=13$ with $R_e = 0.98$ in the case of unbiased erasure ($\eta_e=1/2$, dashed lines) and fully biased erasure ($\eta_e=\infty$, solid lines). The threshold error rate increases from $p_{\text{th}} = 4.3\%$ to $8.2\%$.
{The error bars represent the 95\% confidence interval of the logical error rate, determined from the logical error count and the total number of samples.}
(c) Threshold error rates under different erasure bias for several $R_e$ (from bottom to top, $R_e = 0.5,0.8,0.9,0.98,0.99,1$)}  
\label{fig:threshold_re_etae}
\end{figure}

The analysis above describes infinitely biased erasures arising from the decay of the Rydberg state. However, other mechanisms can also give rise to erasure errors that need to be considered separately.

First, Refs. \cite{jandura2022,fromonteil2022} derived gate protocols that convert Doppler shifts and slow amplitude noise into leakage into the Rydberg state, which can be detected as an erasure error. Since only atoms in $\ket{1}$ are coupled into the Rydberg state in these protocols, as well, the erasure errors should also be strongly biased.

Second, decays out of the qubit subspace $^3P_0$ from Raman scattering from the dipole trap or the finite lifetime of this state also become erasures \cite{Wu2022eras}. However, these are equally likely to occur from either qubit state, and are therefore better represented by the depolarizing channel. The same situation is likely to arise for single-qubit rotations driven by Raman transitions, as in Ref. \cite{jenkins2022yb}.

Finally, erasures in the CZ gates themselves are not infinitely biased if there is a finite probability to excite to (and decay from) the Rydberg state from $\ket{0}$. While this is strongly suppressed in typical hyperfine qubits in alkali atoms because of the large splitting associated with the ground state hyperfine coupling ($\approx 6.8\,$GHz in $^{87}$Rb) \cite{saffman2010quantum}, it is a more significant effect in nuclear spin qubits in alkaline earth atoms because of the smaller splittings. In $^{171}$Yb, the excitation probability of $\ket{0}$ is suppressed by the Zeeman splitting in the Rydberg state \cite{ma2022}. For the $^3S_1$ state, the average occupation of the Rydberg state during a gate when starting from $\ket{0}$ is approximately $p_0 = (\hbar\Omega)^2 / (g_F \mu_B B)^2$, with $g_F = 3/2$ {$g_F = 4/3$ (for $F=3/2$)}. For reasonable values of $\Omega \approx 5\,$MHz, $p_0$ can be suppressed below $10^{-3}$ in a magnetic field of $XX$ Gauss.

{Due to possible erasure from state $\ket{0}$ followed by atom replacement or reinitialization using state $\ket{1}$, the erased qubit can also experience Pauli $\{X,Y\}$ errors.}
To study the effect of finite erasure bias in our model, we introduce the following phenomenological model. When an erasure error occurs during a CZ gate (with probability $p_e$), the state of each atom after recovery is modeled by selecting a Pauli operator from $\{I,Z\}$ with probability $p_I = p_Z \equiv p_a$, or from $\{X,Y\}$ with probability $p_X=p_Y\equiv p_b$, where $2(p_a+p_b)=1$. Defining the erasure bias $\eta_e = p_a/2p_b$, we arrive at the thresholds summarized in Fig.~\ref{fig:threshold_re_etae}(c). In the experimentally relevant range of $R_e$, we see significant improvement in the threshold already for $\eta_e=10$, with saturation effects visible above $\eta_e=100$.

\section{Circuit-based QEC}

Here we describe in more detail the circuit-level simulation of the XZZX surface code. We use a square unrotated XZZX lattice with a code distance $d$ [see Fig.~\ref{fig:threshold_re_etae}(a) for an example of $d=4$].
In {\it step 1}, we initialize all qubits in $\ket{+}$ and then do a round of noiseless stabilizer measurement to initialize the logical-qubit.
In {\it step 2}, we implement $d$ rounds of noisy stabilizer measurement, considering only errors on two-qubit gates.
We flip a coin after every two-qubit gate to include an erasure error with probability $p_e=pR_e$. If an erasure error does not occur, a Pauli error with probability $p_p=p(1-R_e)$ will be introduced.
In {\it step 3}, we do a round of noiseless stabilizer measurement, which can be considered as corresponding to readout of the logical-qubit state. 

Steps 1-3 are implemented using a stabilizer simulator, Stim \cite{Gidney2021_stim,gidney2021honeycomb}.
The stabilizer measurement results give the error syndromes.
Stim can also generate a weighted decoding graph by analysing the effect of individual errors, determining which stabilizers and which logical observables the error flips.
Decoding the error syndromes using the MWPM decoder is done with PyMatching \cite{Higgott2022_pymatching}.

We study three different error models: (i) unbiased erasure, (ii) biased erasure with native gates, and (iii) biased erasure with bias-preserving CX gates.
In case (i) the erasure errors are drawn uniformly at random from $\{I,X,Y,Z\}^{\otimes2}$. In case (ii), for infinitely biased erasure, the erasure errors of CZ gates are drawn uniformly at random from $\{I,Z\}^{\otimes2}$, i.e, each qubit experiences a Pauli operator chosen uniformly at random from $\{I,Z\}$. For CX gates, the erasure errors are drawn from $\{I,Z\}\otimes\{I,X\}$ since CX gates are implemented using CZ gates conjugated by H gates on the target qubit and are thus not bias-preserving. When considering erasure with finite bias $\eta$, following the discussion in Sec.\,\ref{subsec:finite_bias}, the erasure error of a CZ gate is simulated by letting each qubit experience Pauli operators $\{I,Z,X,Y\}$ with probabilities $p_I=p_Z=\eta/(1+2\eta)$ and $p_X=p_Y=0.5/(1+2\eta)$.
In case (iii), for infinitely biased erasure, erasure errors of both CZ and CX gates are drawn uniformly at random from $\{I,Z\}^{\otimes2}$.
In all cases, the remaining Pauli errors (not erasures) are assumed to be unbiased, i.e., drawn uniformly at random from $\{I,X,Y,Z\}^{\otimes2}\backslash \{I\otimes I\}$.
{For all cases, the threshold is determined from the intersection of the curves of the logical error rate versus physical error rate for  $d=9, 13$. We verify the obtained thresholds for $R_e=0.98$ cases in Table I using finite size scaling for $d=9, 11, 13$. As an aside,  the intersection of the $d=9, 11, 13$ curves is similarly used to determine the thresholds in the hybrid-fusion protocol unless otherwise noted.}

We note that since all qubits start in $\ket{+}$, after the initialization in step 1, the logical qubit would be in an eigenstate of the logical X operator (with the eigenvalue being $+1$). Therefore, the derived logical error rate is effectively the logical Z error rate. One can also start with all qubits in state $\ket{1}$ instead of $\ket{+}$ to get the logical X error rate, and
evaluate separate thresholds for logical Pauli-X and Pauli-Z errors (similar to what was mentioned in Ref.\,\cite{Bonilla2021_xzzx}).
In cases (i) and (ii), the two thresholds should be the same since the probabilities of X and Z errors on individual data qubits are equal. This is consistent with our simulations.
In case (iii), since Z errors dominate X errors on all qubits, we evaluate the threshold using the logical Z error rate (note that despite differing error rates, the discrepancy between the two logical thresholds is small).

The step 2 described above, following the procedures in \cite{Wu2022eras}, is used when deriving thresholds corresponding to $R_e=0.98$ (predicted for $^{171}$Yb atom qubits) for comparison with the result in \cite{Wu2022eras}.
In other cases of the circuit-based QEC studied in this work, 
we obtain the thresholds with a slightly modified version of introducing the errors in step 2.
When adding Pauli errors to a two-qubit gate that does not have an erasure, we change the error probability from $p_p$ to $p_p/(1-p_e)$ to represent conditioning on having no erasure.
The corresponding effect on the threshold is small as one can expect from $1-p_e\approx1$. For instance, for $R_e=0.98$ listed in Table 1 in the main text, in the case of infinite erasure bias with native gates, the threshold changes from $8.2\%$ to $8.1\%$. In the cases of unbiased erasure and infinite erasure bias with bias-preserving gates, the thresholds $4.3\%$ and $9.0\%$ have negligible changes.

\section{Fusion circuit}

Here we provide a more detailed discussion of the fusion circuit introduced in the main text. Consider two qubits (labelled $i,j$) which need to be fused and which may be entangled with other qubits (collectively labelled as $R$). We cam write the state of this system as
{
\begin{align}
\ket{\psi}=a\ket{0}_i\ket{0}_j\ket{A}_R+b\ket{1}_i\ket{0}_j\ket{B}_R+c\ket{0}_i\ket{1}_j\ket{C}_R+d\ket{1}_i\ket{1}_j\ket{D}_R
\end{align}}
with $|a|^2+|b|^2+|c|^2+|d|^2=1$. After fusion, $i,j$ are unentangled from the rest of the system and a new state of $R$ is created conditioned on the outcomes $m_\mathrm{XX}$ and $m_\mathrm{ZZ}$ of the $X \otimes X$ and $Z \otimes Z$ measurements respectively.
\begin{equation}
\begin{aligned}
\ket{\psi}_{R,m_\mathrm{XX}=0,m_\mathrm{ZZ}=0}&\propto a\ket{A}_R+d\ket{D}_R\\
\ket{\psi}_{R,m_\mathrm{XX}=1,m_\mathrm{ZZ}=0}&\propto a\ket{A}_R-d\ket{D}_R\\
\ket{\psi}_{R,m_\mathrm{XX}=0,m_\mathrm{ZZ}=1}&\propto b\ket{B}_R+c\ket{C}_R\\
\ket{\psi}_{R,m_\mathrm{XX}=1,m_\mathrm{ZZ}=1}&\propto b\ket{B}_R-c\ket{C}_R
\end{aligned}
\end{equation}
For convenience we have dropped the normalization constants from the above equations. We now explain how this transformation is achieved with the fusion circuit.

In the fusion circuit, a CZ gate is applied between each of the two fusion qubits and the ancilla. In the absence of erasures, which we refer to as {\it case 1}, the ancilla is measured in the $X$ basis, which projects the fusion qubits in an eigenstate of $Z\otimes Z$ with eigenvalue given by the outcome of the ancilla measurement $m_\mathrm{a}$. Thus after the ancilla measurement, the state $\ket{\psi}$ reduces to
\begin{equation}
\begin{aligned}
\ket{\psi}_{m_\mathrm{ZZ}=m_\mathrm{a}=0}&\propto a\ket{0}_i\ket{0}_j\ket{A}_R+d\ket{1}_i\ket{1}_j\ket{D}_R\\
\ket{\psi}_{m_\mathrm{ZZ}=m_\mathrm{a}=1}&\propto b\ket{1}_i\ket{0}_j\ket{B}_R+c\ket{0}_i\ket{1}_j\ket{C}_R
\end{aligned}
\end{equation}
Next we measure $X\otimes I$, $I\otimes X$ with measurement outcomes $m_{i},\;m_{j}$ respectively. After this the fusion qubits are unentangled from $R$ the state of which is $\ket{\psi}_{R,m_\mathrm{XX}=m_{i}\oplus m_{j},m_\mathrm{ZZ}=m_\mathrm{a}}$ as desired. Thus in the absence of errors, the circuit indeed performs the correct fusion operation. 

Next we consider the situation when an erasure is detected. If only the ancilla is erased then each fusion qubit is measured independently in the $Z$ basis with the measurement outcomes $m'_{i},\; m'_{j}$. This gives $m_{ZZ}=m'_{i}\oplus m'_{j}$. In the biased erasure model, the qubit can only be erased if it started from the $\ket{1}$ state. So if one of the fusion qubits is erased then it is effectively measured in the $Z$ basis with measurement outcome $1$. Thus, if only one of the fusion qubits is erased then we measure the other qubit in $Z$ basis to obtain the outcome $m$, so that $m_\mathrm{ZZ}=m\oplus 1$. If both qubits are erased then we know that $m_{ZZ}=0$.

\section{Error correction with planar arrays of qubits}
\label{Sec:fusion_overhead}

\begin{figure*}
\includegraphics[width=\textwidth]{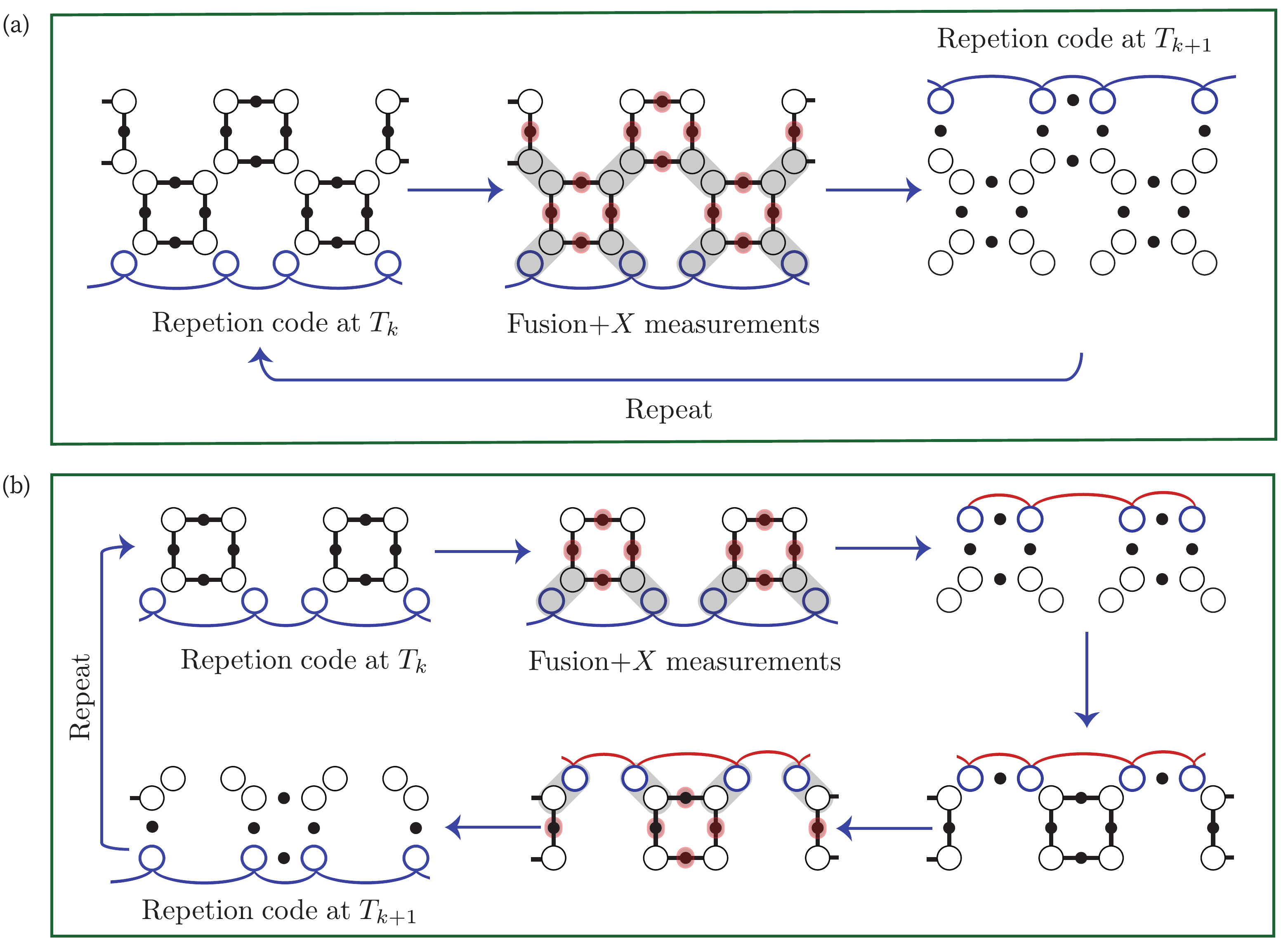}
\caption{(a) Foliated repetition code implemented with just two linear chains of resource states. Note that, due to the structure of our resource states, each stabilizer is measured twice per cycle as shown in this figure. (b) Qubit-efficient implementation of the foliated repetition code. Space is traded off for time/number of steps it takes to measure one round of stabilizers.}
\label{fig:RepCode1}
\end{figure*}

Figure~\ref{fig:RepCode1}(a) shows how a foliated repetition code can be implemented with just two linear chains of resource states. The repetition code is teleported back-and-forth between the qubits marked in blue. We start with repetition code and bring in post-selected resource states. Next we perform the fusion and single qubit $X$ measurements on the two arrays marked with grey and red boxes respectively. This effectively measures all the stabilizers of the repetition code once and teleports the repetition code to the top blue row of qubits. This process can be repeated depending on the desired number of rounds of error correction.

A more qubit-efficient procedure is shown in Fig.~\ref{fig:RepCode1}(b). Here we use only a single array of resource state at a given time, but the array needs to be reinitialized twice per-round of stabilizer measurement. 

\begin{figure*}
\includegraphics[width=0.75\textwidth]{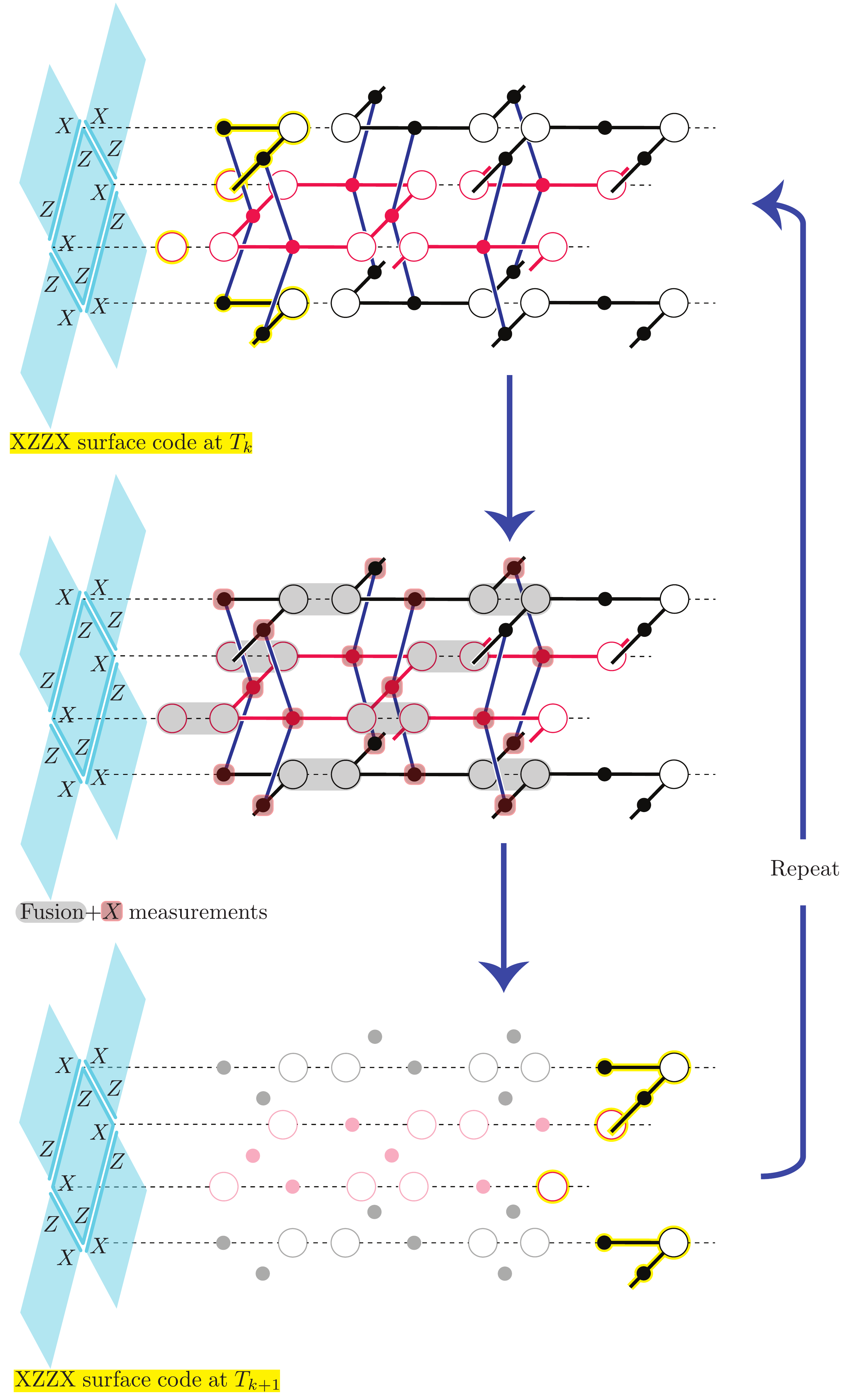}
\caption{XZZX cluster state implemented with planar layers of resource states}
\label{fig:SC}
\end{figure*}

Figure~\ref{fig:SC} shows how the XZZX surface code can be implemented with planar arrays of resource states. The figure focuses on a single stabilzer plaquette for clarity. The code is teleported back-and-forth between the qubits marked in red and black that are placed at the corners of the plaquette. We start with the XZZX code and bring in post-selected resource states. Next we perform the fusion and single qubit $X$ measurements which effectively measures all the stabilizers of the XZZX code once and teleports it to a new plane on the right. This process can be repeated depending on the desired number of rounds of error correction.

\section{Overhead comparison between the circuit-based and fusion-based approaches}

\begin{table}[h]
    \centering
    \begin{tabular*}{\linewidth}{@{\extracolsep{\fill}}lllllll}
    \hline 
    \hline \\[-0.3cm]
    & &$\#$ data & $\#$ ancilla  & $\#$CZ  & $\#$ Meas. &CZ    \\
    & & &   &  &  & depth \\
    \hline
    \\
    \multirow{2}{*}{Rep. code}
     &Fusion  &$d$   &$5d-1$  &$6d-2$   &$5d-1$  &4  \\ \\
      &Fusion (reduced)  &$d$   &$O( 2.5d )$   & $6d-2$   &$5d-1$  &8  \\ \\
    \hline
    \\
    \multirow{5}{*}{Surface  code} & Circuit (rotated) &$d_z^2$ &$d_z^2-1$\footnote{This assumes using different ancilla qubits for different stabilizers, and can be reduced to $(d_z-1)^2$ by reusing some ancilla qubits near the boundary (without qubit shuttling) for different stabilizers \cite{Tomita2014}.} &$4(d_z^2-1)$ &$d_z^2-1$ &4     \\ \\
     &Fusion  &$2d_zd_x-d_x-d_z+1$  & ${O( 11d_zd_x)}$  &${ O( 16d_zd_x)}$  &  $O(10d_zd_x)$&4  \\ \\
      &Fusion (reduced) &$2d_zd_x-d_x-d_z+1$  & ${O (6d_zd_x)}$  &${O( 16 d_z d_x)}$  &  $ O( 10d_zd_x )$&8  \\ \\

    \hline
    \hline
    \end{tabular*}
    \caption{
    {An overhead budget, up to leading order, for the discussed error correction approaches. This budget is calculated per one round of stabilizer measurement. The estimates account for, by labeled column: (1) number of data qubits, (2) number of ancilla qubits, (3) total number of CZ gates, (4) single-qubit measurements (\# Meas.), (5) number of CZ gate layers per round (CZ depth). The `\#CZ' and `CZ depth' in the fusion architecture account for the CZ gates used in resource state generation, fusions, and inter-layer CZs, and also assume successful post-selection of resource states.  For the surface code, we consider a rotated square lattice in the circuit-based approach, while in the fusion-based case, a rectangular lattice is examined due to symmetry preservation of the XZZX code under biased noise even without bias-preserving CX gates.}
    }
    \label{tab:overhead_comparison}
\end{table}

{We present in Table~\ref{tab:overhead_comparison} a resource cost estimate of the two fusion-based approaches for the repetition code, and a comparison of the overhead between the fusion-based and circuit-based approaches for the XZZX code. In the hybrid-fusion approach for the XZZX code, symmetry preservation allows the usage of a rectangular code with $d_\mathrm{x},d_\mathrm{z}$ where $d_\mathrm{x} \neq d_\mathrm{z}$. Here, we use both the previously introduced 8-ring resource states along with 3-body and 5-body states at the boundaries of the code. The estimates in the table account for the bulk of the code; dependent on the choice of starting foliation step there are boundary effects (linear in $d_x,d_z$) that are discarded for this analysis.
Note that the circuit-based approach has to use the square surface code with $d_\mathrm{x}=d_\mathrm{z}$, where we examine the rotated-code configuration for reduced qubit overhead compared to the unrotated code. We note that the threshold error rates presented in the paper are derived using unrotated square codes.}

{We observe that fusion requires significantly more ancilla qubits than circuit-based QEC, $\sim6.5$ times more for a square surface code. This overhead is cut in half by trading space for time with the reduced circuits (\emph{i.e.}, Fig.~\ref{fig:RepCode1}b). However, the fact that the fusion approach maintains the erasure bias allows a rectangular code with $d_x \ll d_z$. This can result in lower qubit counts: for example, if $d_x=3$, the reduced fusion circuit uses fewer qubits than the square code when $d_z \geq 13$.}

{However, we also note that the number of data qubits needed to store a logical state is the same in every approach. In neutral atom qubits, the rate of idle errors is much lower than the rate of gate errors. Therefore, one can envision that for many logical circuit schedules, only a fraction of the logical qubits are undergoing error correction at any point in time. In that case, ancilla qubits for QEC may not be a significant portion of the total qubit budget, and the rectangular codes enabled by the fusion circuit may result in an overhead that is lower by a factor of $d_z/2d_x$.
Finally, we note that the number of CZ gates needed for one QEC round is also larger in the fusion-based approach, but that their influence on the logical error rate is already accounted for in the threshold simulations (i.e, the threshold is higher for the fusion QEC despite the larger number of CZ gates). Furthermore, these gates can be highly parallelized, which is reflected in the small increase in the number of CZ gate rounds.}

\section{Decoding graph for fusion-based construction at $R_e = 1$}
Starting from $8$-ring resource states, the cluster state is built using two primary operations: fusions within 2D planes, and CZ gates between planes. Biased erasures form the dominant error channel for both operations. In the following discussion we set $R_e = 1$, and so Pauli errors are absent. 

\begin{figure*}
\includegraphics[width=\textwidth]{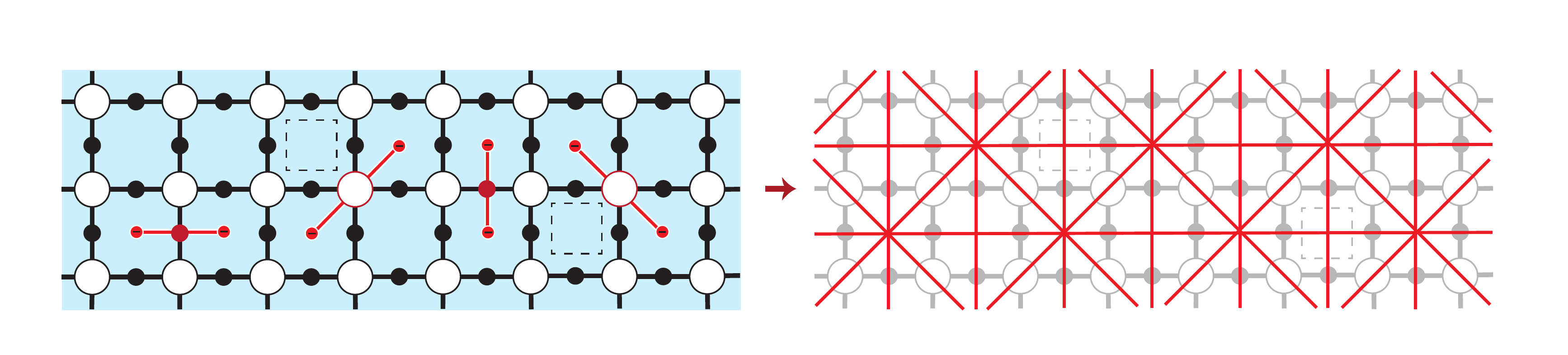}
\caption{(left) Possible error syndromes arising from biased erasures during CZ gates and fusions between $Z$ type qubits for a single plane of the XZZX cluster state. (right) The resultant syndrome graph used for decoding, highlighted in red, forms a union-jack lattice.}
\label{fig:decodinggraph}
\end{figure*}

Fusion failure on a $Z$-type qubit leads to a correlated $Z \otimes Z$ error on its two $X$-type neighbours in the resource state. This correlated error creates $(-1)$ stabilizer outcomes in unit cells diagonally displaced from one another within this plane. An erasure during a CZ gate between two $X$-type qubits leads to a correlated $Z$ error on both qubits. Since these two qubits create pairs of syndromes in unconnected planes, we consider these two errors independently. The resultant decoding graph is shown in Fig.~\ref{fig:decodinggraph}, where the former resource states have been highlighted using dashed lines.

This graph is commonly referred to as the union-jack lattice, and is the dual of the $(4, 8^2)$ tiling of the plane. The percolation threshold $p_{th}$ for this graph is naively $0.3232$~\cite{parviainen2007estimation}, however, because errors from two gates contribute to every edge in the decoding graph, we have $2 p_{th} - p_{th}^2 = 0.3232$, with the quadratic term appearing because we do not perform the CZ gate on erasure detection at the first gate in both fusions and inter-layer CZs on qubits. Consequently, we obtain a theoretical threshold of $p_{th} = 17.7\%$ in terms of the error rate per gate. We recover this value by performing  numerical simulations in the large-size limit. 


\end{document}